# Improving Wheatstone Bridge Sensitivity with Computational Simulations and Bayesian Optimization

Yong Zhou *, Ze-yan Peng, Yan Xiao, Wen-mei Guo and Guan-xin Yao *

Department of Physics, Anhui Normal University, Wuhu 241000, People's Republic of China

E-mail: yong.zhou@mail.ahnu.edu.cn, gx1019@mail.ahnu.edu.cn

**Abstract**

The Wheatstone bridge experiment is fundamental for precise measurement of electrical resistance, holding significant value in both undergraduate physics education and real-life scientific research. This study reimagines the experiment by integrating computational simulation with traditional methods, enhancing its educational and practical value. By analyzing key factors such as internal resistances of the galvanometer and power supply and optimizing resistor configurations, we demonstrate pathways to maximize sensitivity. A Bayesian optimization-based software tool was also developed to automate sensitivity calculations, guiding optimal component selection. This approach bridges theoretical concepts and experimental applications, equipping students with valuable skills in both experimental and computational aspects of physics and preparing students for modern scientific challenges.

Keywords: Wheatstone bridge; sensitivity; simulation; optimization

## 1. Introduction

    The Wheatstone bridge [1,2], comprising four resistors and a galvanometer, was initially proposed by Samuel Christie in 1833 and later popularized by Sir Charles Wheatstone [3]. It is a fundamental circuit used for precise measurement of electrical resistance, holding significant value in both undergraduate teaching of physics experiments and real-life scientific research.

    In undergraduate classes of physical experiments, the Wheatstone bridge serves as a well-regarded comprehensive learning platform that strengthens students' foundational knowledge of electrical circuits, develops their practical lab skills, and introduces them to modern optimization techniques. For example, the Wheatstone bridge is a simple yet powerful circuit for understanding the principles of electrical circuits, including concepts of resistance, current, and voltage [4]. The experiment provides hands-on experience with electrical components such as resistors, galvanometers, and power supplies, reinforcing students' understanding of theoretical concepts [5,6]. Moreover, by performing the Wheatstone bridge experiment, students learn about the importance of precision in measurements and the sensitivity of circuits to small changes in resistance [7].

    Also, the basic working principles underlying the Wheatstone bridge have been widely applied in diverse scientific research. For example, Schopfer and Poirier employed the Wheatstone bridge to examine the universality of the quantum Hall effect [8,9]. Davis and coworkers proposed and theoretically modeled a plasmonic "ac Wheatstone bridge," showing its

potential to detect single molecules [10]. Maedler et al. lithographically fabricated a miniature Wheatstone bridge consisting of four silicon nanowires as arms and reported an enhanced signal-to-noise ratio for pH detection [11]. Li, Liberal, and Engheta extended the Wheatstone bridge with metamaterial-inspired optical nanocircuitry, showing that the high accuracy and simplicity of the Wheatstone bridge may also be exploited in nanophotonic applications such as nanoparticle characterization [12]. Deng et al. designed a novel flexible pressure sensor with a Wheatstone bridge structure, showing improved temperature stability and sensitivity [13]. Also, full Wheatstone-bridge-based sensors are the most useful in giant magnetoresistive (GMR) applications [14]. Li and coworkers developed a microfluidic device to quantify a single cell's mechanical properties by integrating micropipette aspiration to mimic the classical Wheatstone bridge circuit [15]. Recently, fully quantum Wheatstone bridges were proposed [16,17], which find their possible use in fields such as sensing and metrology using near-term quantum devices. Consequently, the Wheatstone bridge's relevance across multiple scientific and engineering disciplines continues to underscore its importance in undergraduate education, preparing students for advanced studies and professional careers in technical fields.

Optimizing the sensitivity of the Wheatstone bridge is fundamental to enhancing its effectiveness in detecting small resistance changes, which is especially important in applications requiring precise and reliable measurements, such as material science, biomedical sensing, and nanotechnology. In an undergraduate physics experiment, high sensitivity is critical for helping students observe and understand subtle variations in electrical resistance, providing a deeper insight into measurement accuracy and precision. Students gain practical circuit design and optimization skills by learning to adjust the circuit parameters to maximize sensitivity, bridging theoretical concepts with hands-on experimentation. This experience not only strengthens their grasp of basic electrical principles but also introduces them to methods of experimental refinement that are essential in scientific and engineering research. However, conventional experimental setups may limit students' ability to explore the full range of parameters affecting bridge sensitivity. Manually adjusting each component in the bridge to optimize sensitivity can be time-consuming, error-prone, and may not always lead to optimal values.

Integrating computer-based tools in educational laboratories has become increasingly popular [18–25], enabling students to engage with simulations and automated calculations that enhance their understanding of complex concepts [26–28]. As a result, there is a need to integrate computational methods that allow students to efficiently explore and analyze the bridge's performance under different conditions. Recent advances in computational techniques, such as Bayesian optimization [29], provide further powerful frameworks for optimizing experimental setups by automatically identifying parameter values that maximize sensitivity. Applying these principles to the Wheatstone bridge makes it possible to develop software that assists students in selecting optimal resistor values, considering practical conditions such as internal resistances of the power supply and galvanometer.

This paper introduces a pedagogical approach to the undergraduate Wheatstone bridge experiment, integrating computational simulation with traditional methods. The influences of factors like internal resistances and resistor configurations on sensitivity are demonstrated, providing deeper insights into optimizing circuit performance. One software named WBOpt was developed using Bayesian optimization principles, allowing students to determine the bridge configuration for maximum sensitivity automatically. The structure of this paper is as follows. Section 2 outlines the basic principles of the Wheatstone bridge experiment and describes our simulation methods. Section 3 presents numerical results relevant to sensitivity simulation and optimization. A summary concludes the article.

## 2. Principles and Methods

### 2.1 Experimental Principles

For the integrity of this paper, we first summarize the basic principles of the Wheatstone bridge experiment. Typically, the bridge circuit is arranged with four resistors in a diamond configuration: two resistors in each of two parallel branches, upper and lower, respectively. The working principle of the Wheatstone bridge relies on the condition of balance, where no current flows through the central detector, e.g., a galvanometer, indicating that the potential difference across it is zero.

As depicted in Fig. 1, in a Wheatstone bridge, two known resistors, $R_3$ and $R_4$, are placed in one branch of the bridge, while a known resistor, $R_2$, and an unknown resistor $R_x$ (the resistance to be measured) are placed in the other branch. These branches are connected across a power source, creating a potential difference. A galvanometer, $G$, connects the midpoints of the two branches (i.e., the "bridge"), allowing for the detection of any current flow between them [4]. In the daily teaching process, we also say that each branch is split into two "arms," and the corresponding $R_2/R_3$ and $R_x/R_4$ arms are named the ratio and comparison arms, respectively.

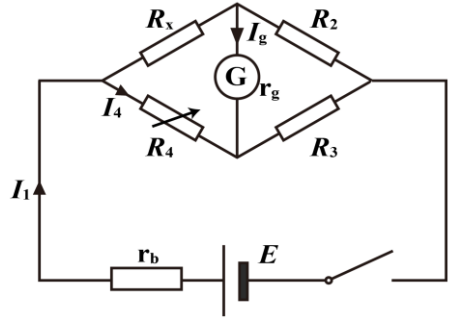

**Figure 1.** Scheme of a typical setup of the Wheatstone bridge experiment.

The bridge reaches a "balanced" state when the ratio of the resistances in one branch is equal to the ratio in the other branch [30],

$$\frac{R_x}{R_4} = \frac{R_2}{R_3}. \tag{1}$$

Under this condition, the voltage drops across each parallel branch are equal, resulting in no potential difference across the central galvanometer. Thus, no current flows through the galvanometer ($I_g = 0$), indicating a balanced bridge.

The Wheatstone bridge can determine the fixed, yet unknown, resistance $R_x$. Here, $R_2$, $R_3$, and $R_4$ can be chosen as standard resistors of known resistance. Precisely, the resistance of $R_4$ is adjusted until the balance condition is fulfilled. In this case, the unknown resistance $R_x$ can be calculated from Equation (1) as

$$R_x = \frac{R_2}{R_3} R_4. \tag{2}$$

Since zero current detection can be achieved with reasonably high accuracy using a galvanometer, $R_x$ can be measured with high accuracy, given that $R_2$, $R_3$, and $R_4$ are known to be precise.

When the requirement of the Equation (1) is not satisfied, a certain current is observed across the bridge. In the Wheatstone bridge experiment, "sensitivity" refers to the responsiveness of the bridge output current to small changes in one of the resistances ($R_4$ here) in the circuit, which is especially important in precision measurements, as the sensitivity determines the bridge's ability to detect minor variations in resistance.

Several definitions for the sensitivity of the Wheatstone bridge were proposed and adopted in the literature, such as the current, voltage, and power ones [7]. For example, Weiss and Maisel employed the dimensionless [31] or normalized [32] sensitivity, expressing the unbalanced voltage as a fraction of the source voltage. In our case, with a constant source voltage of 3.0 V and a pre-selected "bridge ratio" $M = R_2/R_3$, we chose to define the bridge sensitivity as the ratio of the perturbation in current $I_g$ to the relative change in adjustable resistance $R_4$,

$$S = \frac{\Delta I_g}{\Delta R_4 / R_4}. \tag{3}$$

## 2.2 Simulation Methods

As the basis of computer-based simulations and optimizations of bridge sensitivity, we start with a mathematical analysis of the whole circuit using Kirchhoff's laws. By examining Fig. 1, the following equations are derived accordingly,

$$\begin{cases} I_1 \cdot r_b + I_4 \cdot (R_4 + R_3) + I_g \cdot R_3 = E \\ I_1 \cdot R_x - I_4 \cdot (R_4 + R_x) + I_g \cdot r_g = 0 \\ I_1 \cdot R_2 - I_4 \cdot (R_2 + R_3) - I_g \cdot (R_3 + R_2 + r_g) = 0 \end{cases} \tag{4}$$

Equivalently, the Equation (4) can also be written in its matrix form,

$$\begin{bmatrix} r_b & R_4+R_3 & R_3 \\ R_x & -R_x-R_4 & r_g \\ R_2 & -R_2-R_3 & -R_3-R_2-r_g \end{bmatrix} \begin{bmatrix} I_1 \\ I_4 \\ I_g \end{bmatrix} = \begin{bmatrix} E \\ 0 \\ 0 \end{bmatrix}. \tag{5}$$

After defining two determinants, D and $D_g$, as in the Equation (6) and using Cramer's rule, current flows through the galvanometer, $I_g$, can be computed as $I_g = D_g/D$, which applies to both balanced and unbalanced cases of the Wheatstone bridge experiment.

$$D = \begin{vmatrix} r_b & R_4+R_3 & R_3 \\ R_x & -R_x-R_4 & r_g \\ R_2 & -R_2-R_3 & -R_3-R_2-r_g \end{vmatrix}, D_g = \begin{vmatrix} r_b & R_4+R_3 & E \\ R_x & -R_x-R_4 & 0 \\ R_2 & -R_2-R_3 & 0 \end{vmatrix} \tag{6}$$

Computer-based simulations in studying the Wheatstone bridge experiment provide a controlled, versatile, and efficient approach for learning, analysis, and optimization. They are precious for exploring intricate concepts like sensitivity and balance, offering insights that complement hands-on experiments in a time-effective and resource-saving manner. To this end, we have designed and implemented a light-weighted Python program in which two vital functions were defined:

```python
def IG_func(Rx,R2,R3,R4,rb,rg,E):
    Dg=E*(R2*R4-Rx*R3)
    D_matrix=np.array([[rb, R3+R4, R3],
                    [Rx,-R4-Rx, rg],
                    [R2,-R2-R3,-R2-R3-rg]])
    D=np.linalg.det(D_matrix)
    Ig=Dg/D
    return Ig

def sensitivity_func(Rx,R2,R3,R4,rb,rg,E,h):
    Ig_plus_h =IG_func(Rx,R2,R3,R4+h,rb,rg,E)
    Ig_minus_h=IG_func(Rx,R2,R3,R4-h,rb,rg,E)
    derivative=(Ig_plus_h-Ig_minus_h)/(2*h)
    Sl=R4*derivative
    return Sl

def objective(params):
    R2,R3=params
    R4=Rx*R3/R2
    return sensitivity_func(Rx,R2,R3,R4,rb,rg,E,0.0001)

#Bayesian optimization
space=[(0.01,Rx),(0.01,Rx)]
result=gp_minimize(objective,space,n_calls=150,random_state=None)
```

**Box 1**. Code snippet of core functions and their application in sensitivity optimization.

- `IG_func` calculates the current through a galvanometer in a Wheatstone bridge circuit, $I_g$. The corresponding parameters are described as follows. $R_x$, $R_2$, $R_3$, and $R_4$ are resistors in different branches of the Wheatstone bridge, and $r_g$ is the internal resistance of the galvanometer. $E$ and $r_b$ are the voltage and internal resistance of the power supply. The function returns the value of $I_g$, the current through the galvanometer.
- `sensitivity_func` calculates the sensitivity of the current through a galvanometer in a Wheatstone bridge setup concerning a change in one of the resistances, specifically $R_4$. In addition to those parameters used in `IG_func`, one more quantity named $h$ with a default value of 0.0001 Ω, a small increment representing the change in resistance for calculating the derivative is also taken as the input of this function. Herein, the derivative is computed using the central-difference scheme.

Following that, the sensitivity_function is encapsulated into a target function named `objective`, which computes the sensitivity of a Wheatstone bridge circuit based on the values of resistors $R_2$ and $R_3$, those in the ratio arms. Later on, the optimization routine, `gp_minimize` from the *scikit-optimize* library, is utilized to perform the Bayesian optimization to find the optimal values of $R_2$ and $R_3$, aiming at maximizing the sensitivity function within specified bounds. A code snippet is listed in the Box 1 as an example.

## 3. Results and Discussion

As highlighted in the Introduction section of this paper, the sensitivity of the Wheatstone bridge experiment plays a vital role in ensuring accurate, reliable, and precise measurements. Utilizing the computer-based simulation methods described in the last section, we systematically investigated various possible influencing factors on the bridge sensitivity, such as the internal resistances of the galvanometer and power supply and the ratio of the bridge arm resistances.

First, we explore the effect of the galvanometer's internal resistance ($r_g$) on sensitivity since it directly determines the current flowing across the bridge under unbalanced conditions. Specifically, two values for the internal resistance of the galvanometer (100 Ω and 1000 Ω) were considered. The power supply voltage is fixed at 3.0 V throughout this paper, which aligns with our class teaching. An ideal power supply with zero internal resistance is assumed in this part to ease analysis. The resistance to be measured, $R_x$, is 500 Ω, which lies in the medium-value application scope of the Wheatstone bridge. Various combinations of bridge ratio M and pre-selected resistance $R_2$ in one of the ratio arms were investigated, and the resistance $R_3$ in another was determined accordingly. The resistance $R_4$ in the comparison arm is computed based on the working principle given by Equation (1).

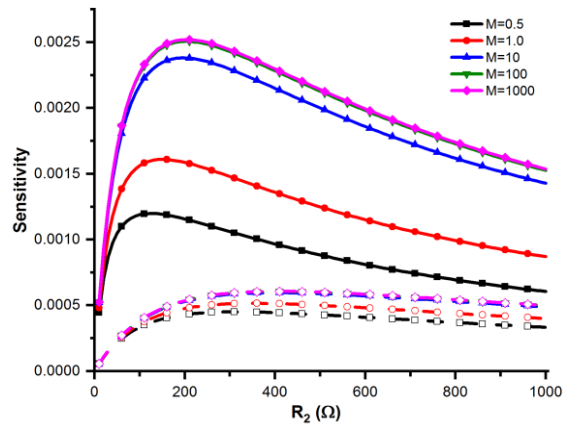

**Figure 2**. Sensitivity as a function of $R_2$ with various bridge ratios (M). Two typical values for the internal resistance of the galvanometer ($r_g$, 100 Ω, and 1000 Ω as solid and dashed lines, respectively) were considered.

Figure 2 depicts the sensitivity as a function of $R_2$ for various bridge ratios (M). Interestingly, for both galvanometer resistances (100 Ω and 1000 Ω), the sensitivity initially increases with $R_2$, reaches a peak, and then decreases. This behavior indicates that the Wheatstone bridge has an optimal $R_2$ value for each configuration where the sensitivity is maximized. Beyond this point, further increases in $R_2$ reduce sensitivity, likely due to diminishing current changes detected by the galvanometer.

As can be seen, the sensitivity curves show significant differences under varied conditions of galvanometer internal resistance $r_g$. The peak sensitivities for each bridge ratio M for $r_g$ = 100 Ω are significantly higher than $r_g$ = 1000 Ω. For example, with M = 1000, the peak sensitivity reaches about 0.0025 for $r_g$ = 100 Ω but only around 0.0005 for $r_g$ = 1000 Ω. Additionally, the $r_g$ = 100 Ω curves maintain a higher sensitivity over a broader range of $R_2$ values than those for $r_g$ = 1000 Ω. Thus, lower internal resistance in the galvanometer improves peak sensitivity and leads to a high sensitivity range, which is readily understood since lowering $r_g$ allows more current to flow through the galvanometer branch under a given resistance imbalance and the resulting voltage difference. This analysis provides a basis for the choice of bridge galvanometers. In other words, to ensure high sensitivity, the internal resistance of the galvanometer should generally be low, typically within the range of tens of Ohms.

Also, Fig. 2 indicates that the ratio of resistances in the bridge is a critical factor for tuning the circuit's responsiveness to changes in $R_4$. Within each set of $r_g$ values, increasing the bridge ratio M (from 0.5 to 1000) leads to higher sensitivity. For instance, the sensitivity curve for M=1000 with $r_g$ = 100 Ω reaches a maximum of around 0.0025, while the curve for M = 0.5 peaks at around 0.0012. Physically, this makes sense because a larger M amplifies the imbalance effect for a given deviation in resistance, making the bridge more sensitive to resistance variations. This characteristic is precious in practical applications where the Wheatstone bridge is used to detect small changes in resistance, such as in strain gauges and temperature sensors. By optimizing M, the bridge can be tailored for maximum sensitivity in these applications.

Considering the power supply's internal resistance allows for a more accurate and reliable analysis of the Wheatstone bridge experiment. It ensures the stability of the voltage across the bridge arms, prevents misinterpretation of results, and contributes to consistent calibration. Figure 3 shows the maximal sensitivity of a Wheatstone bridge as a function of the bridge ratio M under various values of internal resistance of the power supply, denoted as $r_b$. The maximal sensitivities were obtained from curves in analog to those shown in Fig. 2. Here, the internal resistance of the galvanometer $r_g$ has been fixed at its actual value of 40 Ω, as measured in our laboratory. For each curve in Fig. 3, the sensitivity increases sharply as the bridge ratio M rises from zero, then gradually reaches a peak, followed by a slight decrease or plateau as M continues to increase. This observation indicates that there is an optimal bridge ratio that maximizes sensitivity, beyond which further increases in M do not enhance sensitivity and may even reduce it slightly.

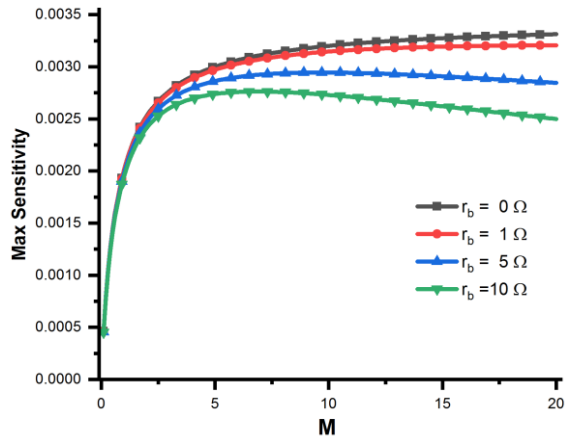

**Figure 3**. Maximal sensitivity as a function of the bridge ratio (*M*) for various internal resistances of the power supply (*r*$_b$).

Moreover, Fig. 3 also highlights the significant impact of the power supply's internal resistance on the sensitivity. The curve with $r_b$ = 0 Ω (ideal power supply) achieves the highest sensitivity across the entire range of M considered, while the sensitivity decreases progressively with higher $r_b$ values. Quantitatively, for $r_b$ = 0, the peak sensitivity reaches approximately 0.0033. As $r_b$ increases to 10 Ω, the peak sensitivity drops to around 0.0025. This difference indicates that a non-zero internal resistance in the power supply reduces the overall bridge sensitivity, making the system less responsive to small resistance changes. Besides that, the optimal M for peak sensitivity shifts slightly depending on $r_b$. For lower $r_b$ values ($r_b$ = 0 Ω or $r_b$ = 1 Ω), the sensitivity peaks above M = 15. The peak occurs at a lower M value between 5 and 10 for higher $r_b$ values ($r_b$ = 5 Ω or $r_b$ = 10 Ω), suggesting that higher internal resistance reduces the effective bridge ratio required for maximal sensitivity.

Figure 4 shows a heatmap plot of the bridge sensitivity as a function of resistances $R_2$ and $R_3$ with a non-zero internal resistance of the power supply ($r_b$) of 5 Ω, which provides a more comprehensive perspective on the behavior of the bridge circuit. The color gradient represents the bridge sensitivity, with the scale on the right indicating the sensitivity values from blue (low) to red (high). According to the color scale, the sensitivity ranges from 0.00005 to approximately 0.00295. This wide range of values underscores how dramatically the sensitivity of the bridge can vary depending on the chosen resistance values.

The plot exhibits a clear color gradient, ranging from blue on the left to red on the right. The blue areas (low sensitivity) are concentrated where $R_2$ values are low, particularly around $R_2$ < 50 Ω. Red areas (high sensitivity) dominate as $R_2$ and $R_3$ increase, suggesting that higher values of these resistances generally enhance sensitivity. Specifically, as $R_2$ increases, sensitivity consistently rises, while increasing $R_3$ has a similar but slightly less pronounced effect. This trend implies an optimal operating region regarding $R_2$ and $R_3$ values, where sensitivity is maximized, which is crucial for applications requiring high measurement precision.

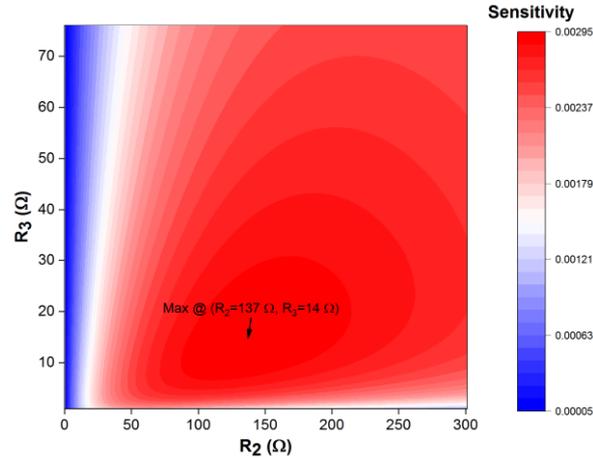

**Figure 4**. Heatmap plot of bridge sensitivity as a function of $R_2$ and $R_3$, with non-zero internal resistance of the power supply ($r_b$).

The highest sensitivity is observed in the central lower regions of the plot, especially around $R_2 = 137\ \Omega$ and $R_3 = 14\ \Omega$. This "hot spot" area indicates that, despite the limitation on the bridge's performance due to non-zero internal resistance of the galvanometer ($r_g$) and power supply ($r_b$), careful adjustment of $R_2$ and $R_3$ can compensate to some extent, pushing the sensitivity into higher regions. In practical applications, such a plot can guide the selection of resistance values for specific sensitivity requirements, balancing precision and the impact of the internal resistances.

Carefully chosen values for $R_2$ and $R_3$ and thus the resulting bridge ratio $M$ can streamline calibration processes, enhance sensitivity, and ultimately improve the reliability of measurement outcomes. However, achieving this precise adjustment can be challenging due to the vast space of experimental parameters. Therefore, developing computer-based software to optimize these parameters is crucial.

Figure 5 displays a snapshot of the user-friendly software, WBOpt [33], we developed to optimize the sensitivity of a Wheatstone bridge using Bayesian optimization principles. This approach minimizes manual adjustments and ensures sensitivity calculations quickly converge to optimal values with several iterations. With essential parameters—such as the estimated resistance $R_x$, power supply voltage and internal resistance (E and $r_b$), and the galvanometer's internal resistance $r_g$—inputted, this tool automates sensitivity calculations, guiding users in selecting the optimal component values to enhance the experiment's sensitivity, accuracy, and measurement reliability, while significantly reducing the reliance on manual trial and error. Such software is precious for both educational and scientific purposes.

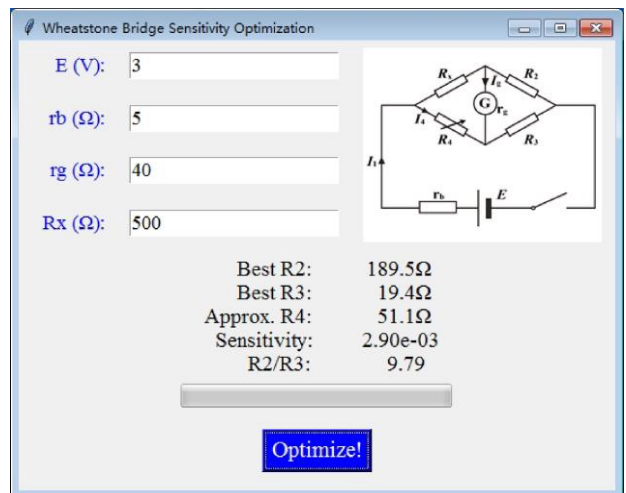

**Figure 5**. Snapshot of the software searching for the optimal experiment setup.

## 4. Conclusions

This study presents an integrated approach to the Wheatstone bridge experiment, combining computational simulations with traditional experimental methods. By systematically analyzing factors such as the galvanometer and power supply's internal resistances and resistor configurations, we highlighted their impact on the sensitivity of the Wheatstone bridge. Our findings reveal that reducing internal resistances and carefully selecting bridge ratios are crucial to achieving maximum sensitivity, making detecting minor resistance variations with higher accuracy possible. To enhance the pedagogical value, we developed a Bayesian optimization-based software tool that allows students to input circuit parameters and compute optimal configurations efficiently. This computational assistance minimizes manual adjustments and deepens students' understanding of the relationship between circuit parameters and measurement sensitivity. Our methodology enriches the learning experience in undergraduate physics education and bridges the gap between theory and application by emphasizing both practical experimentation and computational modeling. Furthermore, the study demonstrates the continued relevance of the Wheatstone bridge as a versatile educational tool, providing foundational skills essential for both experimental physics and interdisciplinary applications in engineering and technology.

## Acknowledgements

This research was funded by the Anhui Provincial Quality Engineering Project of Higher Education (Grant Nos. 2020jyxm1980, 2019mooc051), the Anhui Provincial Leading Talents Project, and the Anhui Provincial Innovation Project of Returnees from Overseas Studies.

## Data availability statement

All data that support the findings of this study are included within the article (and any supplementary files).